# Integral order photonic RF and microwave signal processors based on soliton crystal Kerr micro-combs

Mengxi Tan,[1] Xingyuan Xu,[2] David J. Moss[1]

[1]Optical Sciences Centre, Swinburne University of Technology, Hawthorn, VIC 3122, Australia.
[2]Department of Electrical and Computer Systems Engineering, Monash University, Clayton, 3800 VIC, Australia.

E-mail: dmoss@swin.edu.au

**Abstract**

Soliton crystal micro-combs are powerful tools as sources of multiple wavelength channels for radio frequency (RF) signal processing. They offer a compact device footprint, large numbers of wavelengths, very high versatility, and wide Nyquist bandwidths. Here, we demonstrate integral order RF signal processing functions based on a soliton crystal micro-comb, including a Hilbert transformer and first- to third-order differentiators. We compare and contrast results achieved and the tradeoffs involved with varying comb spacing, tap design methods, as well as shaping methods.

**1. Introduction**

RF signal processing functions, including the Hilbert transform and differentiation, are building blocks of advanced RF applications such as radar systems, single sideband modulators, measurement systems, speech processing, signal sampling, and communications [1-42]. Although the electronic digital-domain tools that are widely employed enable versatile and flexible signal processing functions, they are subject to the bandwidth bottleneck of analog-to-digital convertors [4], and thus face challenges in processing wideband signals.

Photonic RF techniques [1-3] have attracted great interest during the past two decades with their capability of providing ultra-high bandwidths, low transmission loss, and strong immunity to electromagnetic interference. Many approaches to photonic RF signal processing have been proposed that take advantage of the coherence of the RF imprinted optical signals – thereby inducing optical interference. These coherent approaches map the response of optical filters, implemented through optical resonators or nonlinear effects, onto the RF domain [7-12]. As such, the ultimate performance of the RF filters largely depends on the optical filters. State-of-art demonstrations of coherent photonic RF filters include those that use integrated micro-ring resonators, with Q factors of > 1 million, as well as techniques that employ on-chip (waveguide-based) stimulated Brillouin scattering [10-12]. Both of these approaches have their unique advantages - the former uses passive devices and so can achieve very low power consumption, while Brillouin scattering can achieve a much higher frequency selectivity, reaching a 3 dB bandwidth resolution as low as 32 MHz.

Coherent approaches generally focus on narrow-band applications where the frequency range of concern is narrow and the focus is on frequency selectivity, and where the filters are generally band-pass or band-stop in nature. In contrast, incoherent approaches that employ transversal filtering structures can achieve a very diverse range of functions with a much wider frequency range, such as Hilbert transforms and differentiators. The transversal structure originates from the classic digital finite impulse response filter, where the transfer function is achieved by weighting, delaying and summing the input signals. Unlike digital approaches that operate under von-Neumann protocols, photonic implementations achieve the entire process through analog photonics, where the weighting, delaying and summing happens physically at the location of the signals, instead of reading and writing back-and-forth from memory.

To achieve the transversal structure optically, four steps are required. First, the input RF signals are replicated, or multicast, onto multiple wavelengths simultaneously using wavelengths supplied from either multiple single wavelength, or single multiple wavelength, sources. Next, the replicated signals are assigned different weights for each wavelength and then the composite signal is progressively delayed where each wavelength is incrementally delayed relative to the next. Finally, the weighted replicas are summed together by photodetecting the entire signal. The underpinning principle to this process is to physically achieve multiple parallel channels where each channel carries and processes one replica of the RF signal. In addition to wavelength multiplexing techniques, this can also be accomplished with spatial multiplexing, such using an array of fibre delay lines to spatially achieve the required parallelism. Although this is straightforward to implement, it suffers from severe tradeoffs between the number of channels and overall footprint and cost. Exploiting the wavelength dimension is a much more elegant approach since it makes much better use of the wide optical bandwidth of over the 10 THz that the telecommunications C-band offers, and thus is more compact. However, traditional approaches to generating multiple optical wavelengths have been based on discrete laser arrays, [6-9] and these face limitations in terms of a large footprint, relatively high cost, and challenges in terms of accurate control of the wavelength spacing.

Optical frequency combs - equally spaced optical frequency lines - are a powerful approach to implementing incoherent photonic RF filters since they can provide a large number of wavelength channels with equal frequency spacings, and in a compact scheme. Among the many traditional methods of achieving optical frequency combs, electro-optic (EO) techniques have probably experienced the widest use for RF photonics. By simultaneously driving cascaded EO modulators with a high-frequency RF source, a large number of comb lines can be generated, and these have been the basis of many powerful functions. However, EO combs are not without challenges. On the one hand, they generally have a small Nyquist zone (half of the frequency spacing), limited by the RF source. On the other hand, the employed bulky optical and RF devices are challenging to be monolithically integrated. As such, to overcome the hurdles of size, reliability and cost-effectiveness of bulky photonic RF systems, integrated frequency combs would represent a highly attractive approach.

Integrated Kerr optical frequency combs [47-76], or micro-combs, that originate via optical parametric oscillation in monolithic micro-ring resonators (MRRs), have recently come into focus as a fundamentally new and powerful tool due to their ability to provide highly coherent multiple wavelength channels in integrated form, from a single source. They offer a much higher number of wavelengths than typically is available through EO combs, together with a wide range of comb spacings (free spectral range (FSR)) including ultra-large FSRs, as well as greatly reduced footprint and complexity. Micro-combs have enabled many fundamental breakthroughs [50] including ultrahigh capacity communications [77-79], neural networks [80-82], complex quantum state generation [83-97] and much more. In particular, micro-combs have proven to be very powerful tools for a wide range of RF applications such as optical true time delays [31], transversal filters [34, 38], signal processors [29, 32], channelizers [37] and others [15, 18, 26-28, 36, 39-41]. They have greatly enhanced the performance of RF signal processors in terms of the resolution (for coherent systems) and operation bandwidth (for incoherent systems).

In one of the first reports of using micro-combs for RF signal processing, we demonstrated a Hilbert transformer based on a transversal filter that employed up to 20 taps, or wavelengths. [36] This was based on a 200 GHz FSR spaced micro-comb source that operated in a semi-coherent mode that did not feature solitons. Nonetheless, this provided a low enough noise comb source to enable very attractive performance, achieving a bandwidth of over 5 octaves in the RF domain. Subsequently, [15] we demonstrated $1^{st}$ $2^{nd}$ and $3^{rd}$ order integral differentiators based on the same 200 GHz source, achieving high RF performance with bandwidths of over 26 GHz, as well as a range of RF spectral filters including

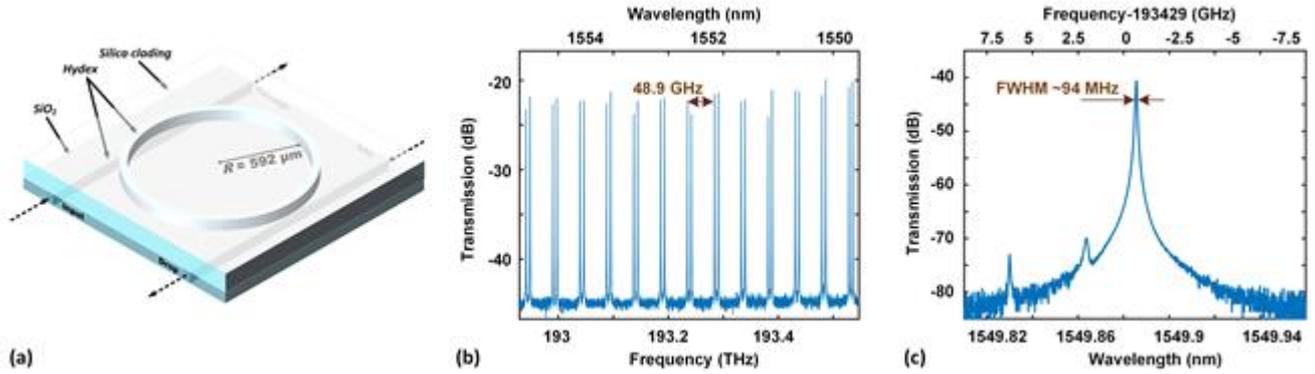

Fig. 1. (a) Schematic of the micro-ring resonator. (b) Drop-port transmission spectrum of the integrated MRR with a span of 5 nm, showing an optical free spectral range of 48.9 GHz. (c) A resonance at 193.429 THz with a full width at half maximum (FWHM) of ~94 MHz, corresponding to a quality factor of ~$2\times10^6$.

bandpass, tunable bandpass and gain equalizing filters [32, 33].

Recently, a powerful category of micro-combs — soliton crystals — has been reported [59, 76, 98]. It features ultra-low intensity noise states and straightforward generation methods via adiabatic pump wavelength sweeping. Soliton crystals are unique solutions to the parametric dynamics governed by the Lugiato-Lefever equation. They are tightly packaged solitons circulating along the ring cavity, stabilized by a background wave generated by a mode-crossing. Due to their much higher intra-cavity intensity compared with the single-soliton states of DKS solitons, thermal effects that typically occur during the transition from chaotic to coherent soliton states are negligible, thus alleviating the need for complex pump sweeping methods.

We have exploited soliton crystal states generated in record low FSR (49 GHz) micro-ring resonators (MRRs), thus generating a record large number of wavelengths, or taps, to achieve a wide range of high performance RF signal processing functions. These include RF filters [35], true time delays [30], RF integration [42], fractional Hilbert transforms [27], fractional differentiation [41], phase-encoded signal generation [26], arbitrary waveform generation [43], filters realized by bandwidth scaling [38], and RF channelizers [44] and much more [99-110].

In this work, we further examine transversal photonic RF signal processors that exploit soliton crystal micro-combs. We demonstrate Hilbert transformers as well as 1$^{st}$, 2$^{nd}$, and 3$^{rd}$ order integral differentiators and explore in detail the trade-offs inherent between using differently spaced soliton crystal micro-combs as well as different numbers of tap weights and design methods. Our study sheds light on the optimum number of taps, while the experimental results agree well with theory, verifying the feasibility of our approach towards the realization of high-performance photonic RF signal

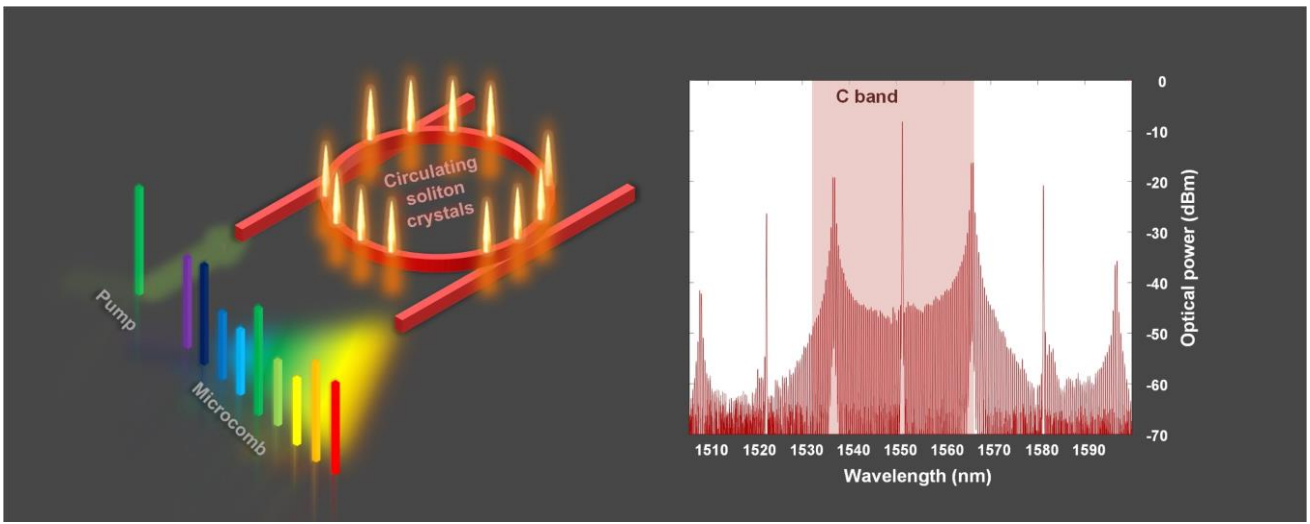

Fig. 2. Schematic illustration of the integrated MRR for generating the Kerr frequency comb and the optical spectrum of the generated soliton crystal combs with a 100-nm span.

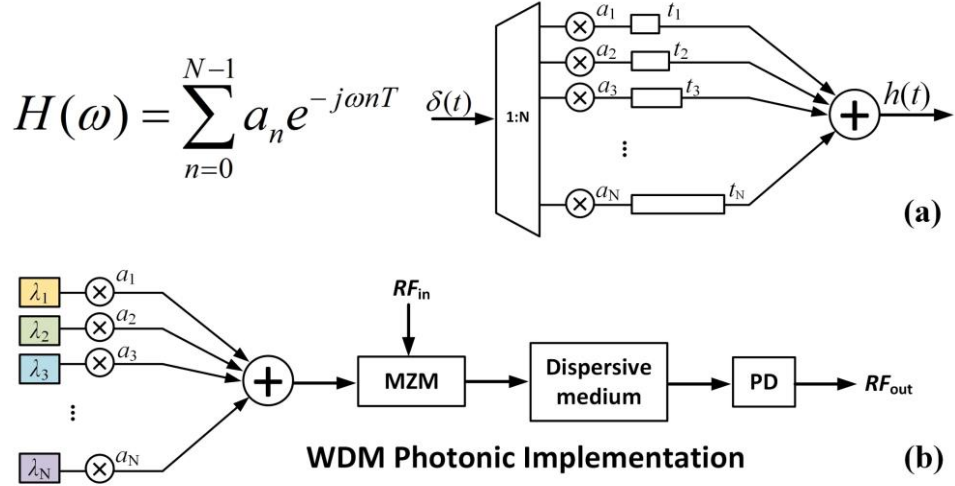

Fig. 3. Conceptual diagram of the transversal structure.

processing with potentially reduced cost, footprint and complexity.

## 2. Operation principle

The generation of micro-combs is a complex process that generally relies on a high nonlinear material refractive index, low linear and nonlinear loss, as well as engineered anomalous dispersion [59-64]. Diverse platforms have been developed for micro-comb generation [58], such as silica, magnesium fluoride, silicon nitride, and doped silica glass. The MRR used to generate soliton crystal micro-combs is shown in Fig. 1 (a). It was fabricated on a high-index doped silica glass platform using CMOS compatible processes. Due to the ultra-low loss of our platform, the MRR features narrow resonance linewidths, corresponding to quality factors as high as 1.5 million, with radii of ~592 µm, which corresponds to a very low FSR of ~0.393 nm (~48.9 GHz) (Fig. 1 (b)) [54-55, 39-40]. First, high-index ($n$ = ~1.7 at 1550 nm) doped silica glass films were deposited using plasma-enhanced chemical vapour deposition, followed by patterning with deep ultraviolet stepper mask photolithography and then etched via reactive ion etching followed by deposition of the upper cladding. The device architecture typically uses a vertical coupling scheme where the gap (approximately 200 nm) can be controlled via film growth – a more accurate approach than lithographic techniques. The advantages of our platform for optical micro-comb generation include ultra-low linear loss (~0.06 dB·cm$^{-1}$), a moderate nonlinear parameter (~233 W$^{-1}$·km$^{-1}$) and, in particular, a negligible nonlinear loss up to extremely high intensities (~25 GW·cm$^{-2}$) [65-76]. After packaging the device with fibre pigtails, the through-port insertion loss was as low as 0.5 dB/facet, assisted by on-chip mode converters.

To generate soliton crystal micro-combs, we amplified the pump power up to 30.5 dBm. When the detuning between the pump wavelength and the cold resonance became small enough, such that the intra-cavity power reached a threshold value, modulation instability (MI) driven oscillation was initiated. Primary combs were thus generated with a spacing determined by the MI gain peak – mainly a function of the intra-cavity power and dispersion. As the detuning was changed further, distinctive 'fingerprint' optical spectra were observed (Fig. 2), similar to what has been reported from

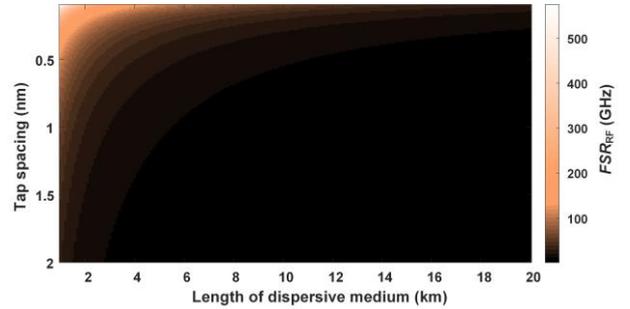

Fig. 4. Free spectral range of the RF transversal signal processor according to the length of fibre and comb spacing. Here we used single mode fibre with the second order dispersion coefficient of $\beta$ = ~17.4 ps/nm/km at 1550 nm for the calculation of $FSR_{RF}$.

spectral interference between tightly packed solitons in a cavity – so-called 'soliton crystals' [55-56]. The second power step jump in the measured intra-cavity power was observed at this point, where the soliton crystal spectra appeared. We found that it was not necessary to achieve any specific state, including either soliton crystals or single soliton states, in order to obtain high performance – only that the chaotic regime [59] should be avoided. Nonetheless, the soliton crystals states provided the lowest noise states of all our micro-combs and have also been used as the basis for a microwave oscillator with low phase-noise [28]. This is important since there is a much wider range of coherent low

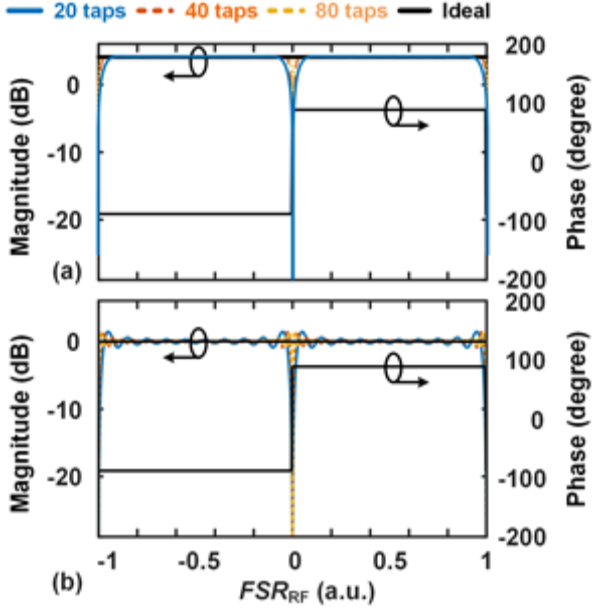

Fig. 5. Theoretical and simulated RF magnitude according to the number of taps and ideal phase response of Hilbert transformer with 90° phase shift. (a) With a hamming window applied. (b) Without window method applied.

RF noise states that are more readily accessible than any specific soliton related state [59].

Figure 3 illustrates the conceptual diagram of the transversal structure. A finite set of delayed and weighted replicas of the input RF signal are produced in the optical domain and then combined upon detection. The transfer function of a general transversal signal processor can be described as

$$H(\omega) = \sum_{n=0}^{N-1} a_n e^{-j\omega nT} \quad (1)$$

where $N$ is the number of taps, $\omega$ is the RF angular frequency, $T$ is the time delay between adjacent taps, and $a_n$ is the tap coefficient of the $n_{th}$ tap, which is the discrete impulse response of the transfer function $F(\omega)$ of the signal processor. The discrete impulse response $a_n$ can be calculated by performing the inverse Fourier transform of the transfer function $F(\omega)$ of the signal processor [11]. The free spectral range of the RF signal processor is determined by $T$, since $FSR_{RF} = 1/T$. As the multi-wavelength optical comb is transmitted through the dispersive medium, the time delay can be expressed as

$$T = D \times L \times \Delta\lambda \quad (2)$$

where $D$ denotes the dispersion coefficient, $L$ denotes the length of the dispersive medium, and $\Delta\lambda$ represents the wavelength spacing of the soliton crystal micro-comb, as shown in Fig. 4, which indicates the potentially broad bandwidth RF signal that the system can process. From Figure 4 we can see the relationship between the wavelength spacing of the comb, the total delay of the fibre, and the resulting RF FSR, or essentially Nyquist zone. The operation bandwidth can be easily adjusted by changing the time delay (i.e., using different delay elements). The maximum operational bandwidth of the transversal signal processor is limited by the comb spacing (i.e., the Nyquist frequency, or half of the comb spacing). Thus, employing a comb shaping method to achieve a larger comb spacing could enlarge the maximum operational bandwidth, although at the expense of providing fewer comb lines/taps across the C-band. Hence, the number of comb lines/taps as well as the comb spacing, are key parameters that determine the performance of the signal processor. We investigate this tradeoff in detail in this paper.

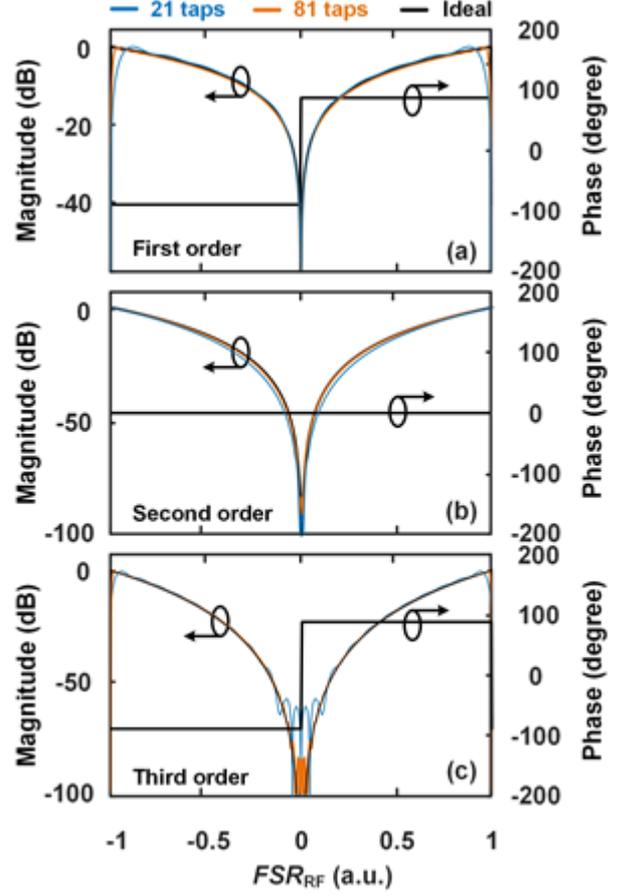

Fig. 6. Theoretical and simulated RF magnitude according to the number of taps and ideal phase response of (a) first-order differentiator. (b) second-order differentiator. (c) third-order differentiator.

Figures 5 and 6 show the theoretically calculated performance of the Hilbert transformer with a 90° phase shift together with the 1st, 2nd and 3rd order integral differentiators in terms of their filter amplitude response, as a function of the number of taps. Note that a Hamming window [11] is applied in Fig. 5 (a) in order to suppress the sidelobes of the Hilbert transformer. As seen in Fig. 7, the theoretical 3 dB bandwidth increases rapidly with the number of taps.

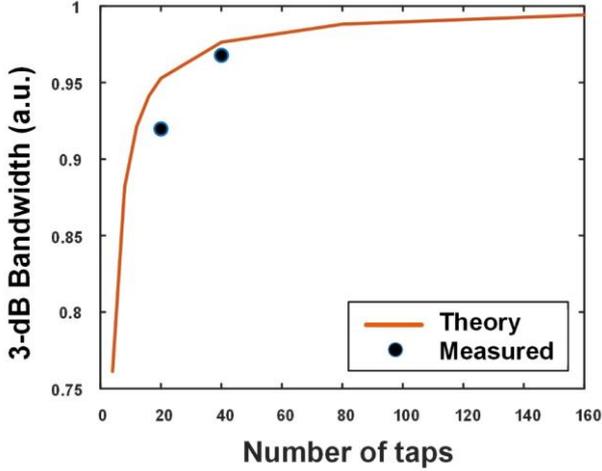

Fig. 7. Simulated and experimental results of 3-dB bandwidth with different taps for a Hilbert transformer with 90° phase shift.

## 3. Experiment

Figure 8 shows the experimental setup of the transversal filter signal processor based on a soliton crystal micro-comb. It consists mainly of two parts - comb generation and flattening followed by the transversal structure. In the first part, the generated soliton crystal micro-comb was spectrally shaped with two WaveShapers to enable a better signal-to-noise ratio as well as a higher shaping accuracy. The first WaveShaper (WS1) was used to pre-flatten the scallop-shaped comb spectrum that is a hallmark of soliton crystal micro-combs. In the second part, the flattened comb lines were modulated by the RF input signal, effectively multicasting the RF signal onto all of the wavelength channels to yield replicas. The RF replicas were then transmitted through a spool of standard SMF ($\beta$ = ~17.4 ps/nm/km) to obtain a progressive time delay between the adjacent wavelengths. Next, the second WaveShaper (WS2) equalized and weighted the power of the comb lines according to the designed tap coefficients. To increase the accuracy, we adopted a real-time feedback control path to read and shape the power of the comb lines accurately. Finally, the weighted and delayed taps were combined and converted back into the RF domain via a high-speed balanced photodetector (Finisar, 43 GHz bandwidth).

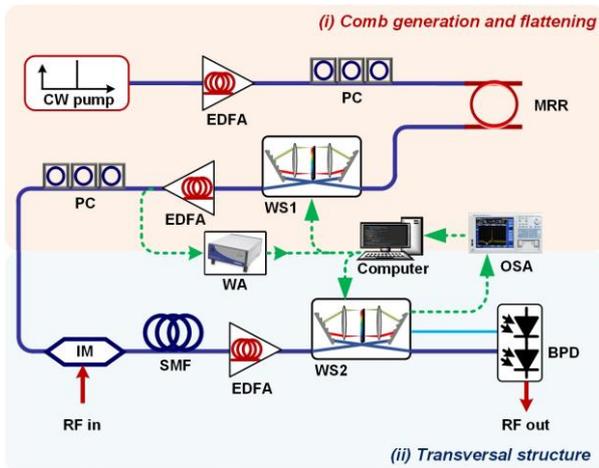

Fig. 8. Experimental set up of RF signal processor based on soliton crystal micro-comb source. CW: Continuously wave. EDFA: Erbium-doped fibre amplifier. PC: Polarization controller. WS: WaveShaper. IM: Intensity modulator. SMF: Single mode fibre. BPD: Balanced photodetector. WA: wave analyzer. OSA: optical spectral analyzer

Figure 9 shows the experimental results for the Hilbert transformer with a 90° phase shift. The shaped optical combs are shown in Figs. 9 (a) (e) (i). A good match between the measured comb lines' power (blue lines for positive, black lines for negative taps) and the calculated ideal tap weights (red dots) was obtained, indicating that the comb lines were successfully shaped. Note that we applied a Hamming window [11] for single-FSR (49 GHz) and 4-FSR (196 GHz) comb spacings when designing the tap coefficients. One can see that with a Hamming window applied, the deviation of the amplitude response from the theoretical results can be improved. Figs. 9 (b) (f) (j) show the measured and simulated amplitude response of the Hilbert transformer using single-FSR, 2-FSR, and 4-FSR comb spacings, respectively. The corresponding phase responses are depicted in Figs. 9 (c) (g) (k). It can be seen that all three configurations exhibit the response expected from the ideal Hilbert transform. The system demonstration for the Hilbert transform with real-time signals consisting of a Gaussian input pulse, generated by an arbitrary waveform generator (AWG, KEYSIGHT M9505A) was also performed, as shown in Figs. 9 (d) (h) (l) (black solid curves). They were recorded by means of a high-speed real-time oscilloscope (KEYSIGHT DSOZ504A Infinium). For comparison, we also depict the ideal Hilbert transform results, as shown in Figs. 9 (d) (h) (l) (blue dashed curves). For the Hilbert transformer with single-FSR, 2-FSR, and 4-FSR comb spacings, the calculated RMSEs between the measured and the ideal curves are ~0.133, ~0.1065, and ~0.0957, respectively. The detailed performance parameters are listed in Table 1.

Figure 10 shows the experimental results for the differentiators with increasing integral orders of 1, 2, and 3. The shaped optical spectra in Figs. 10 (a) (e) (i) (m) (q) (u) show a good match between the measured comb lines' power and the calculated ideal tap weights. Figures. 10 (b) (f) (j) (n) (r) (v) show measured and simulated amplitude responses of the differentiators. The corresponding phase response is depicted in Fig. 10 (c) (g) (k) (o) (s) (w) where it can be seen that all integral differentiators agree well with theory. Here, we use the WaveShaper to programmably shape the combs to simulate MMRs with different FSRs. By essentially artificially adjusting the comb spacing, we effectively obtain a variable operation bandwidth for the differentiator, which is advantageous for diverse requirements of different applications. Here, we normalised the FSR of the RF response to have the unique operational bandwidth for comparing the perfoamance of different processing functions in the same scales. For the 1$^{st}$, 2$^{nd}$, and 3$^{rd}$ order differentiators with a single-FSR (49 GHz) spacing, the calculated RMSEs between the measured and ideal curves are ~0.1111, ~0.1139, ~0.1590, respectively. For the 1$^{st}$, 2$^{nd}$, and 3$^{rd}$ order differentiators with a 4-FSR (196 GHz) spacing, the calculated RMSEs between the measured and ideal curves are ~0.0838, ~0.0570, ~0.1718, respectively. Note that there is some observed imbalance in the time-domain between the positive and negative response to the Gaussian input pulse. This is due to the imbalance of the two ports of the balanced photodetector.

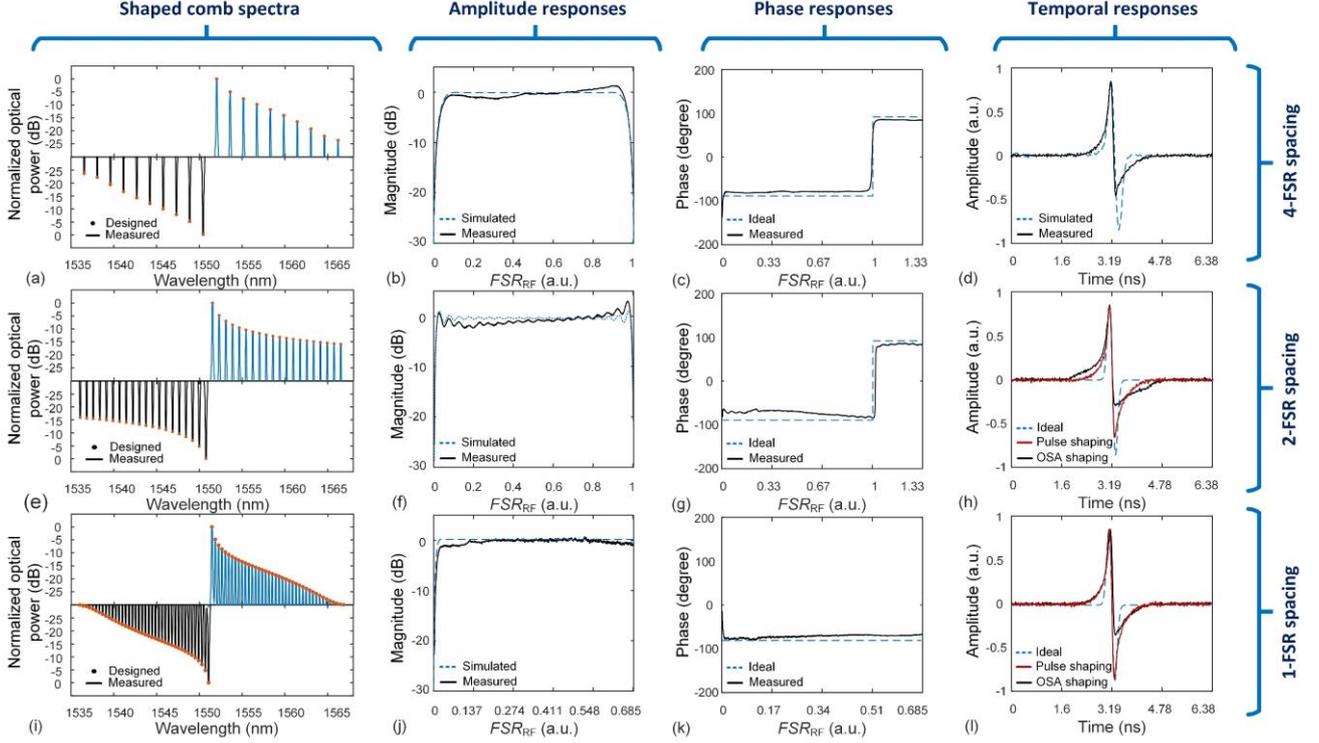

Fig. 9. Simulated and measured 90° Hilbert transformer with varying comb spacing. (a) (e) (i) Shaped optical spectral. (b) (f) (j) Amplitude responses (the |S21| responses measured by a Vector Network Analyzer). (c) (g) (k) Phase responses. (d) (h) (l) Temporal responses measured with a Gaussian pulse input.

In order to reduce the errors mentioned above, for both the Hilbert transformer and the differentiator, we developed a more accurate comb shaping approach, where the error signal of the feedback loop was generated directly by the measured impulse response, instead of the optical power of the comb lines. We then performed the Hilbert transform and differentiation with the same transversal structure as the previous measurements, the results of which are shown in Figs. 9 (h) (I) and Fig. 10 (t). One can see that the imbalance of the response in time domain has been compensated, and the RMSE of time-domain shown in Table 1 has significantly improved.

TABLE I
PERFORMANCE OF OUR TRANSVERSAL SIGNAL PROCESSORS

| Type | Number of taps | Wavelength spacing | Frequency spacing (GHz) | Nyquist zone (GHz) | Octave | Temporal pulse RMSE | |
| --- | --- | --- | --- | --- | --- | --- | --- |
| | | | | | | OSA shaping | Pulse shaping |
| Hilbert transformer | 20 | 4-FSR | 196 | 98 | > 4.5 | ~0.0957 | / |
| Hilbert transformer | 40 | 2-FSR | 98 | 49 | > 6 | ~0.1065 | ~0.0845 |
| Hilbert transformer | 80 | Single-FSR | 49 | 24.5 | / | ~0.1330 | ~0.0782 |
| Differentiator – 1$^{st}$ order | 21 | 4-FSR | 196 | 98 | / | ~0.0838 | / |
| Differentiator – 2$^{nd}$ order | 21 | 4-FSR | 196 | 98 | / | ~0.0570 | / |
| Differentiator – 3$^{rd}$ order | 21 | 4-FSR | 196 | 98 | / | ~0.1718 | / |
| Differentiator – 1$^{st}$ order | 81 | Single-FSR | 49 | 24.5 | / | ~0.1111 | / |
| Differentiator – 2$^{nd}$ order | 81 | Single-FSR | 49 | 24.5 | / | ~0.1139 | ~0.0620 |
| Differentiator – 3$^{rd}$ order | 81 | Single-FSR | 49 | 24.5 | / | ~0.1590 | / |

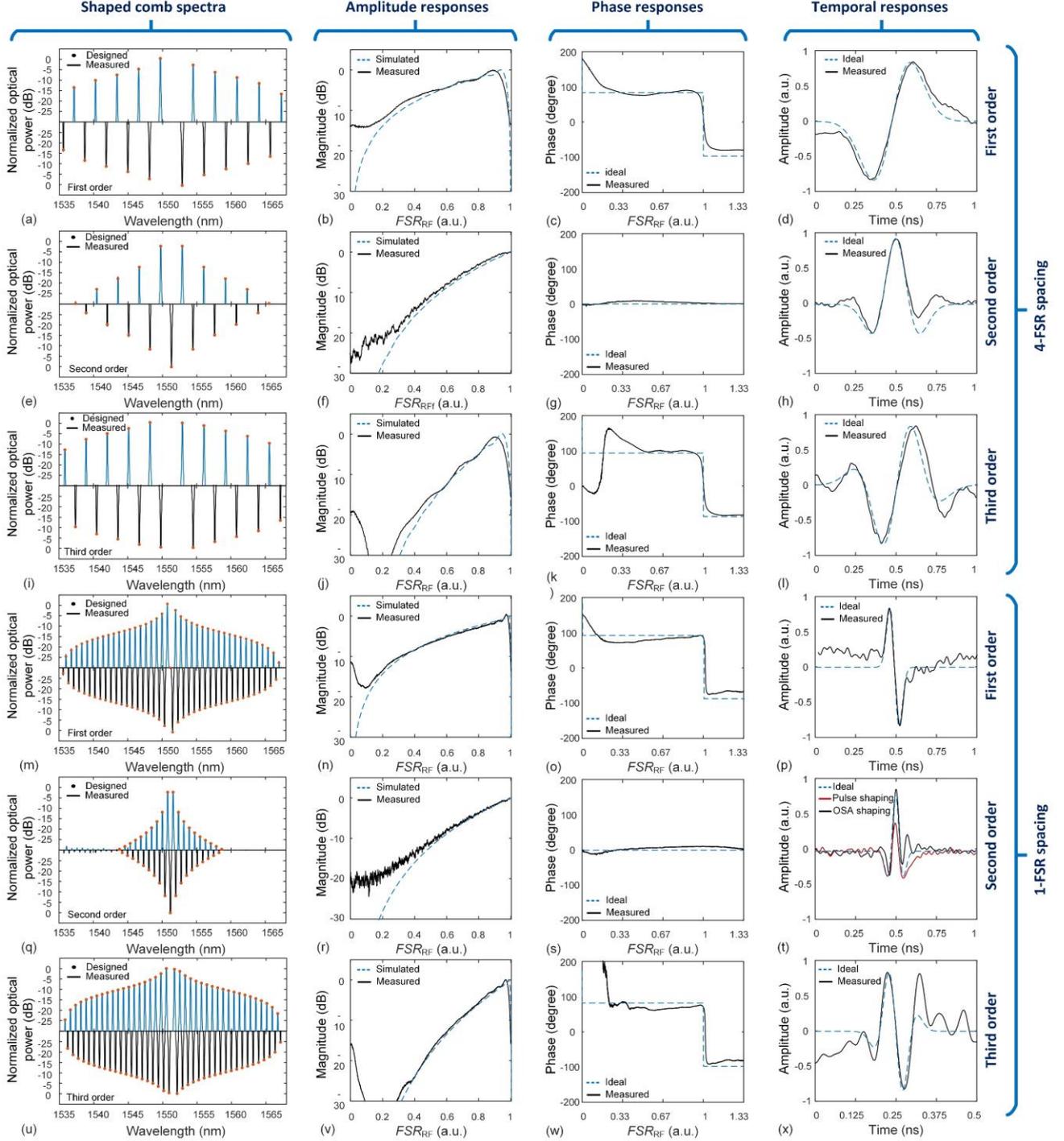

Fig. 10. Simulated and measured first- to third-order differentiators with different comb spacing (single-FSR and 4-FSR). (a) (e) (i) (m) (q) (u) Shaped optical spectral. (b) (f) (j) (n) (r) (v) Amplitude responses. (c) (g) (k) (o) (s) (w) Phase responses. (d) (h) (l) (p) (t) (x) Temporal responses measured with a Gaussian pulse input.

Also note that the greater number of lines supplied by the soliton crystal micro-comb (81 for the 1-FSR spacing) yielded significantly better performance in terms of the spanned number of octaves in the RF domain as well as the RMSE, etc.

On the other hand, the 1-FSR spacing is more limited in operational bandwidth, being restricted to roughly the Nyquist zone of 25 GHz. The 2-FSR spacing and 4-FSR spacing system can reach RF frequencies well beyond what conventional electronic microwave technology can achieve.

Therefore our shaping method gives the flexibility for us to achieve the required system.

## 4. Conclusion

We demonstrate record performance and versatility for soliton crystal micro-comb-based RF signal processing functions by varying wavelength spacing and employing different tap designs and shaping methods. The experimental results agree well with theory, verifying that our soliton crystal micro-comb-based signal processor is a competitive approach towards achieving RF signal processor with broad operation bandwidth, high reconfigurebility, and potentially reduced cost and footprint.

Competing interests: The authors declare no competing interests.

**Acknowledgments**

This work was supported by the Australian Research Council Discovery Projects Program (No. DP150104327). RM acknowledges support by the Natural Sciences and Engineering Research Council of Canada (NSERC) through the Strategic, Discovery and Acceleration Grants Schemes, by the MESI PSR-SIIRI Initiative in Quebec, and by the Canada Research Chair Program. He also acknowledges additional support by the Government of the Russian Federation through the ITMO Fellowship and Professorship Program (grant 074-U 01) and by the 1000 Talents Sichuan Program in China. Brent E. Little was supported by the Strategic Priority Research Program of the Chinese Academy of Sciences, Grant No. XDB24030000.